\documentclass[letterpaper,11pt]{article}
\usepackage{jheppub}
\usepackage{amssymb}
\usepackage{amsbsy}
\usepackage{amsfonts}
\usepackage{amsmath}
\usepackage{dsfont}
\usepackage{epsfig}
\usepackage{psfrag}
\usepackage{subfigure}
\usepackage{graphicx}
\usepackage{epstopdf}

\def\bea{\begin{eqnarray}}
\def\eea{\end{eqnarray}}
\def\beqn{\begin{eqnarray}}
\def\eeqn{\end{eqnarray}}
\def\beq{\begin{equation}}
\def\eeq{\end{equation}}

\def\Dslash{\not{\hbox{\kern-4pt $D$}}}
\def\pslash{\not{\hbox{\kern-4pt $p$}}}

\def\D{\mathcal{D}}
\def\U{\mathcal{U}}

\newcommand{\eins}{\mbox{$1 \hspace{-1.0mm} { \bf l}$}}

\title{UV friendly T-parity in the $\mathbf{SU(6)/Sp(6)}$ little Higgs model.}
\author[a]{Tom Brown}
\emailAdd{gregoire@physics.carleton.ca,claudia.frugiuele@gmail.com, tombrown321@yahoo.ca}
\author[a]{Claudia Frugiuele}
\author[a]{Thomas Gr\'egoire}

\affiliation[a]{Ottawa-Carleton Institute for Physics , Department of Physics,  Carleton University, \\
               1125 Colonel By Drive, Ottawa, Canada, K1S 5B6}
\abstract{Electroweak precision tests put stringent constraints on the parameter space of little Higgs models. Tree-level exchange of TeV scale particles in a generic little Higgs model  produce higher dimensional operators that make contributions to electroweak observables that are typically too large. To avoid this problem a discrete symmetry dubbed T-parity can be introduced to forbid the dangerous couplings. However, it was realized that in simple group models such as the littlest Higgs model, the implementation of T-parity in a UV completion could present some challenges. The situation is analogous to the one in QCD where the pion can easily be defined as being odd under a new $Z_2$ symmetry in the chiral Lagrangian, but this $Z_2$ is not a symmetry of the quark Lagrangian. In this paper we examine the possibility of implementing a T-parity in the low energy $SU(6)/Sp(6)$ model that might be easier to realize in the UV.  In our model, the T-parity acts on the low energy non-linear sigma model field in way which is different to what was originally proposed for the Littlest Higgs, and lead to a different low energy theory. In particular, the Higgs sector of this model is a  inert two Higgs doublets model with an approximate custodial symmetry. We examine the contributions of the various sectors of the model to electroweak precision data, and to the dark matter abundance.
}
\arxivnumber{1012.2060}

\begin{document}
\maketitle

\section{Introduction}
Little Higgs models are constructions where the electroweak symmetry is broken by a scalar Higgs field which is light because it is a pseudo-Goldstone boson \cite{ArkaniHamed:2001nc, ArkaniHamed:2002pa,ArkaniHamed:2002qx,ArkaniHamed:2002qy,Kaplan:2003uc} (see \cite{Kaplan:1983sm,Dugan:1984hq} for an older realization of the Higgs as a pseudo-Goldstone). To realize this idea, first a global symmetry group is spontaneously broken via an unspecified mechanism. This will generate various Goldstone bosons, among which is the Higgs. Then weak couplings are introduced to give the Higgs appropriate gauge, Yukawa and scalar quartic couplings. In order for these couplings not to introduce large radiative corrections to the Higgs mass, they must break the global symmetry only collectively \cite{ArkaniHamed:2002qy}. That is each coupling alone, while breaking a part of the global group, must leave enough of it unbroken such that the Higgs remains an exact massless Goldstone boson. Only when two or more weak couplings are introduced in the theory will the Higgs get a mass, guarantying that it will remain parametrically light. In this way, the Higgs mass can be made lighter than $f$, the Goldstone boson decay constant, by a loop factor and therefore the cutoff of the theory where the Goldstone bosons become strongly coupled can be as high as $\sim 4 \pi f \sim 10$ TeV. Note that this last estimate is an upper bound on the scale of the cutoff as studies of perturbative unitarity in these models \cite{Chang:2003vs} show that new physics, not described in the low energy non-linear sigma model limit,  close to $\sim 4 f$  is needed to restore unitarity.  

The description of the Higgs and its couplings at  low energy, i.e. below $\sim 10$ TeV, can be done using non-linear sigma model fields parametrized by a set of pseudo-Goldstone bosons. Besides the usual Standard Model fields, this description will include extra scalars, extra gauge bosons and extra colored fermions. These new states have  masses near $f \sim 1$ TeV and are responsible for canceling the one loop quadratic divergences to the Higgs mass that occur in the Standard Model. Above $10$ TeV, the derivative couplings of the Goldstone bosons render the theory strongly coupled and a new description, a UV completion, has to be used. 

The main phenomenological problems of little Higgs models come from the fact that the new TeV states can contribute to electroweak precision observables at tree level \cite{Csaki:2002qg,Csaki:2003si,Gregoire:2003kr,Kilic:2003mq, Kilian:2003xt, Barbieri:2004qk,Marandella:2005wd,Han:2005dz}. These contributions  are parametrically of the same size as loop contributions of Standard Model particles or contributions from a strongly coupled sector at the TeV scale such as technicolor. These tree-level contributions are calculable, and can be made small  but in general they limit the parameter space of little Higgs models significantly. A perhaps more elegant solution is to invoke a symmetry principle to forbid the dangerous tree level contributions. The most economical symmetry would be a $Z_2$ symmetry, similar to the $R$-parity that is imposed  in the MSSSM, under which all the new states are odd while the Standard Model particles are even.  In the context of little Higgs, such a parity was defined in various models and dubbed T-parity \cite{Cheng:2003ju,Cheng:2004yc,Low:2004xc}.

In the Littlest Higgs model \cite{ArkaniHamed:2002qy} for example the T-parity acts schematically as \cite{Cheng:2004yc}:
\begin{equation}
\label{eq:tdeflittlest}
\Sigma \rightarrow \Omega \Sigma^{\dagger} \Omega^\dagger \; \;,
\end{equation}
where $\Sigma = \exp(2 i \pi) \Sigma_0$ is a $5 \times 5$ non-linear sigma model field containing the Goldstone bosons that transforms as a symmetric two-index tensor of the global $SU(5)$ and $\Omega$ is an element of $SO(5)$. It was however later realized that this type of $Z_2$ symmetry might not be straightforward to implement in a UV completion \cite{Hill:2007nz,Hill:2007zv}. For example in a strongly coupled UV completion where the non-linear sigma model field is a composite of fermions $\psi_a$ \cite{Katz:2003sn}(with $a$ a $SU(5)$ index):
\begin{equation}
\Sigma_{a b} \sim \psi_a \psi_b \;,
\end{equation}
 T-parity in the UV theory would act as:
\begin{equation}
\psi_a \rightarrow \psi_b^\dagger
\end{equation}
which is not a symmetry of the fermionic kinetic terms. Only in combination with a space-time parity: $\mathbf{x} \rightarrow - \mathbf{x}$, is this a symmetry. The manifestation of this fact in the low energy theory is the presence of  WZW-like  terms that break $T$-parity \cite{Hill:2007nz,Hill:2007zv}(but not the T-parity combine with $\mathbf{x} \rightarrow - \mathbf{x}$). While those terms do not lead to dangerous contributions to electroweak precision measurements, they make the dark matter candidate unstable. 
And while one could impose such a space-time parity on the gauge sector, the situation is more problematic in the fermion sector. Again schematically, T-parity invariant Yukawa couplings require terms of the form:
\begin{equation}
Q_1 \Sigma u_1^c +  Q_2 \Sigma^\dagger u_2^c
\end{equation}
where the two terms are related by T-parity. In a UV completion, these terms could come, for example from 4-Fermi operators:
\begin{equation}
\frac{Q_1^a \psi_a \psi_b {u_1^c}^b + Q_2^a \psi_a^\dagger \psi_b^\dagger {u_2^c}^b}{f^2}
\end{equation}
where, again the two terms need to have the same coefficient. In this case it is hard to see what symmetry of the UV theory could relate the two terms \footnote{In fact, since the Standard Model is not symmetric under $\mathbf{x} \rightarrow -\mathbf{x}$, this symmetry would need to be broken somehow.}.

In this work  we examine the possibilities for redefining T-parity so that it could be integrated without difficulty in a UV completion. We find the $SU(6)/Sp(6)$ little Higgs model \cite{Low:2002ws} particularly amenable to such a redefinition \footnote{This is also one of the two smallest special coset models without a dangerous singlet \cite{Schmaltz:2008vd,Hook:2009kx}} .  Besides implementing T-parity, there exist other challenges in building UV completions of little Higgs models \cite{Katz:2003sn}. In particular, the fermion sector is difficult to obtain through a strongly coupled UV completion without fine-tuning in most little Higgs models \cite{Batra:2007iz} (see \cite{Katz:2003sn} for a possible solution). In addition, in the $SU(6)/Sp(6)$ model, wether or not one can get the correct vaccuum alignment to work in  a strongly coupled UV completion remains an open question \cite{Piai:2004yb}.  The lack of calculability in models where the UV completion is strongly coupled is a major obstacle in obtaining definitive answers to these questions, and we do not attempt to do so in this work. We instead focus on the implementation of T-parity in the low energy theory and on its consequences. We indeed find that the new definition of T-parity that we propose leads to a different phenomenology in the Higgs sector, both at colliders and for cosmology.

 After exposing the leading order gauge and fermionic structure of  a T-symmetric version of the $SU(6)/Sp(6)$ model in sections \ref{sec:model}, we compute various radiative corrections to the Higgs potential in section \ref{sec:radcorr}. We finally discuss the consequences of the model for electroweak precision tests in section \ref{sec:ewpm}  and for dark matter abundance and phenomenology in sections \ref{sec:DM} and \ref{sec:pheno}.

\subsection{T-parity in the Littlest Higgs model}
Before turning to the $SU(6)/Sp(6)$ model we now discuss briefly the challenges associated with building a strongly coupled UV completion to a the Littlest Higgs model with T-parity. To implement T-parity in a Littlest Higgs-like model with an $SU(5)/SO(5)$ global group structure, in a way that does not involve the complex conjugate of the non-linear sigma model field as in \eqref{eq:tdeflittlest} , one could imagine using instead an exchange symmetry \cite{Krohn:2008ye,Freitas:2009jq}.  The field content could be doubled, with two non-linear sigma model fields $\Sigma_A$ and $\Sigma_B$ which are interchanged under T-parity. The Lagrangian for the non-linear sigma model fields would simply be:
\begin{equation}
\mathcal{L}= \left| D_\mu \Sigma_A \right|^2 + \left| D_\mu \Sigma_B \right|^2 \; \;,
\end{equation}
where both transform as $\Sigma_i \rightarrow U \Sigma_i U^T$ under $SU(5)$. The field content is doubled, and contains both odd and even fields. To decouple the unwanted even fields, we would need to write a term of the form:
\begin{equation}
\label{eq:sigAsigB}
c \Lambda^2 f^2 \text{Tr} (\Sigma_A \Sigma_B^\dagger) \;.
\end{equation}
In the limit where $c\sim 1$, the even combination of the Goldstone fields decouple and one is left with the original littlest Higgs model. The main complication for the UV completion is then to generate that term.  For example, lets imagine that the fields $\Sigma_A$ and $\Sigma_B$ are composites of  strongly interacting fermions $\psi_A$ and $\psi_B$ respectively. Both $\psi_A$ and $\psi_B$ are charged under a different strongly interacting $SO(N)$ gauge group and are are fundamental of an $SU(5)$ flavor group. In fact there are now also two different $SU(5)$ global flavor groups, one associated to $\psi_A$, the other to $\psi_B$. To break this enlarged flavor group to the diagonal through a term of the form of (\ref{eq:sigAsigB}),  one could write 4-fermi operators of the form:
\begin{equation}
\label{eq:4fermi}
\text{Tr} \left(\frac{\psi_A \psi_A^T \psi_B^* \psi_B^\dagger}{M^2} \right) \;.
\end{equation}
This would require the introduction of a new scale $M$, close to $f$. Note also that (\ref{eq:4fermi}) needs to be $SU(5)$ invariant. Generating (\ref{eq:sigAsigB}) with a coefficient proportional to $f^4$ instead of $f^2 \Lambda^2$ might be easier as the $SU(5)$ invariance requirement would be softened, and the mass scale $M$, could be above $\Lambda$, in a regime where $\psi_A$ and $\psi_B$ are still weakly coupled. In this case however, the new even fields have masses of order $f \sim 1$ TeV, and one must check that they do not reintroduce large correction to the electroweak precision observables. This could lead to a viable model but the structure of the theory, if not the low energy theory at least the UV completion would need to be enlarged. Note however that the situation is  simpler in a weakly coupled supersymmetric UV completion such as the one built in \cite{Csaki:2008se}.

 \section{The SU(6)/Sp(6) model with T-parity}
 \label{sec:model}
 \subsection{Global symmetries}
 \label{sec:global}
 We now turn to the $SU(6)/Sp(6)$ model\cite{Low:2002ws}, for which we find an alternative to doubling the content of the theory. A successful implementation of a T-parity needs the lead to new gauge bosons that are odd, new scalars that are odd, especially if these new scalars contain a triplet and finally, in the fermion sector, we need the Standard Model fermions to be even.
 
 In the Littlest Higgs model the gauge sector consists of two $SU(2)$ groups: $SU(2)_A$ and $SU(2)_B$, and possibly two $U(1)$ groups. The Standard Model gauge bosons and the new TeV scale gauge bosons are orthogonal linear combinations of $V_A^\mu$ and $V_B^\mu$, the gauge bosons associated with the two groups. To make the Standard Model gauge bosons even and the new TeV scale gauge boson odd, we need a $Z_2$ symmetry that would interchange $V_A^\mu$ and $V_B^\mu$. In a prototype UV completion with strongly coupled fermions (see for example \cite{Katz:2003sn}), this would mean:
 \begin{equation}
 \psi_{2_A} \rightarrow \psi_{2_B} \; \; ,
 \end{equation}
where ${\psi_{2_A}}$ and  $\psi_{2_B}$ are doublet of $SU(2)_A$ and $SU(2)_B$ respectively. However, in a littlest Higgs type model, $\psi_{2_A}$ and $\psi_{2_B}$ have different hypercharge, so this $Z_2$ symmetry would not commute with $U(1)_Y$. If we consider instead the $SU(6)/Sp(6)$ model, where the equivalent fields have hypercharge $0$, this sort of symmetry can be implemented. The spontaneous breaking of the global $SU(6)$ to $Sp(6)$ is in this case parametrized at low energy by the $ 6 \times 6$ non-linear sigma model field:
 \begin{equation}
  \Sigma= e^{ i \Pi_a X_a/f} \  \Sigma_0 \ e^{ i \Pi_a X_a^T/f}=e^{ 2 i \Pi_a X_a/f} \  \Sigma_0,
  \end{equation}
  where 
  \begin{equation}
   \Sigma_0 = f \begin{pmatrix}    0  &  0&  -  \eins_{2 \times 2}  \\ 
0 & \epsilon_{2 \times 2} &0 \\ 
   \eins_{2\times 2}  &0 &   0  \\ 
 \end{pmatrix} ,\
 \label{eq:vacuum}
\end{equation}  
is the vacuum of the theory before electroweak symmetry breaking and each entry in the matrix is itself a $2 \times 2$ matrix with $\epsilon$ the $2 \times 2$ antisymmetric matrix with $\epsilon_{1 2}=-1$. There are $14$ Goldstone bosons, which can be written in matrix form:
\begin{equation}
 \Pi=\begin{pmatrix}  \tilde{\phi}- \frac{\eta}{2} & h_A&h_B& \chi \epsilon \\ 
h_A^{\dagger} & \frac{ \eta}{2} & 0& -h^T_B \\ 
h_B^{\dagger} &0 &\frac{ \eta}{2}  &  h^T_A \\ 
 - \epsilon \chi^{\dagger}& -h_B^* &  h_A^*& \tilde{\phi}^T-\frac{\eta}{2} \\ 
 \end{pmatrix} .
  \label{eq:Pi}
\end{equation}
 where  $\tilde{\phi}$ is an hermitian $2 \times 2$ matrix, $h_A$ and $h_B$ are 2 components column vectors, $\eta$ is a real field and $\chi$ a complex field. Gauge quantum numbers of these fields will be determined by the gauging of various subgroups of the global $SU(6)$ group presented in the next subsection.
   \subsection{Gauge interactions}
   \label{sec:gauge}
 Gauge interactions are introduced as usual by gauging two $SU(2)$ subgroups of the $SU(6)$ global group:
  \begin{equation}
 X_A^a=\begin{pmatrix}   \sigma^a & 0&0 \\ 
0& 0_{2 \times 2}&0 \\ 
 0&0 &0_{2 \times 2} \\ 
 \end{pmatrix},
 \label{eq:q1}
\end{equation}
\begin{equation}
 X_{B}^a=\begin{pmatrix}  0_{2 \times 2}& 0&0 \\ 
0& 0_{2\times 2}&0 \\ 
 0&0 &-  {\sigma^a}^* \\ 
 \end{pmatrix} ,
  \label{eq:q2}
\end{equation}
where $\sigma_a$ are the Pauli matrices with $\text{Tr} ( \sigma_a \sigma_b ) = \delta_{a b}/2$. Only the diagonal subgroup, identified with the Standard Model $SU(2)_L$ is left unbroken by the vacuum, leaving one massless $SU(2)$ triplet of gauge bosons, and one massive one. The gauge couplings associated with $SU(2)_A$ and with $SU(2)_B$ will be equal due to T-parity: $g_A=g_B=g$. We can also establish that $\phi$ is a triplet of $SU(2)_L$, while $h_A$ and $h_B$ are doublets and $\chi$ and $\eta$ are singlets. The hypercharge gauge couplings are introduced by gauging a different subgroup of $Sp(6)$:
\begin{equation}
\label{eq:gaugeY}
Y = \begin{pmatrix}   0_{2 \times 2}& 0&0&0 \\ 
0& 1/2&0&0 \\ 
 0&0 &-1/2&0 \\ 
 0 &0&0&0_{2 \times 2}
 \end{pmatrix}
\end{equation}
with the hypercharge gauge boson.

Implementing T-parity in the fermion sector, which will be presented in the next section, is made easier by the introduction of a third $SU(2)$. This can be seen in the following way: if we introduce a fermion doublet $q_A$ transforming under $SU(2)_A$, T-parity forces the introduction of another doublet transforming under $SU(2)_B$:
\begin{equation}
q_A \rightarrow q_B \; \;.
\end{equation}
The odd combination must be made massive, and an even combination will be massless and identified with a SM fermion doublet. Naively, this can be done by adding a mass term of the form
\begin{equation}
(q_A - q_B) q_C^c \; \;.
\end{equation}
This is invariant under the Standard Model $SU(2)$ but not invariant separately under $SU(2)_A$ and $SU(2)_B$. To make it invariant, we add a third $SU(2)$ gauge group called $SU(2)_C$ , whose associated gauge boson is even under T-parity. We break the resulting $SU(2)^3$ gauge group to its diagonal subgroup by adding two link fields $K_A$ and $K_B$ (\cite{Csaki:2008se,Pappadopulo:2010jx}) that transform as  bifundamental of $SU(2)_A \times SU(2)_C$ and $SU(2)_B \times SU(2)_C$ respectively and get vev's proportional to the identity:
\begin{eqnarray}
K_A &\rightarrow& U_A K_A U_C^\dagger, \; \; \; \; \; \; \left<K_A \right> = \eins_{2 \times 2} ,\\
K_B &\rightarrow& U_B K_B U_C^\dagger ,\; \; \;  \; \; \; \left<K_B \right> = \eins_{2  \times 2},
\end{eqnarray}
where $U_A,U_B$ and $U_C$ are elements of $SU(2)_A,SU(2)_B$ and $SU(2)_C$ respectively. The Lagrangian for the kinetic term of the Goldstone bosons of the theory is given by:
\begin{equation}
\label{eq: kin}
\mathcal{L}_\text{gauge} = \frac{1}{8} f^2 \text{Tr} \left| D_\mu \Sigma \right|^2+ f_k^2 \text{Tr} \left|D_\mu K_A \right|^2 + f_k^2 \text{Tr} \left| D_\mu K_B \right|^2
\end{equation}
where $K_A$ and $K_B$ can be written in term of the Goldstone modes $\phi_A^a$ and $\phi_B^a$ as:
\begin{equation}
K_A = \exp\left(i \phi_A^a \frac{\sigma_a}{f_k}\right) \; \; \; \; K_B = \exp\left(i \phi_B^a \frac{\sigma_a}{f_k} \right) \;.
\end{equation}
 The gauge bosons mass eigenstates and their masses can be derived from equation \eqref{eq: kin}:
 \begin{align}
M^2_{\tilde{V}} &= \frac{1}{2} g^2 ( f^2+ 2 f_k^2) , &
 \tilde{V}^{\mu}&= \frac{( V^{\mu}_A-V^{\mu}_B)}{\sqrt{2}} , \\
M^2_{V_H}&=  f_k^2 (g+ 2g_C^2),  & 
 V^{\mu}_H&= \frac{( g V^{\mu}_A+ g V^{\mu}_B- 2 g_C V^{\mu}_C )}{\sqrt{ 2  g^2+4 g_C^2}}, \\
 M^2_{V_{SM}}&=0, & 
 V^{\mu}_{SM}&= \frac{( g_C V^{\mu}_A+ g_C V^{\mu}_B+g V^{\mu}_C )}{\sqrt{  g^2+ 2 g_C^2}} .
\end{align}
Two of the gauge bosons, $V^{\mu}_\text{SM}$ and $V^{\mu}_H $, are even  and one, $\tilde{V}^{\mu}$, is odd  under T-parity, which acts on the gauge bosons as 
\begin{equation}
V_A^\mu \rightarrow V_B^\mu ,\; \; \; \; \;\; V^\mu_C \rightarrow V^\mu_C \;.
\end{equation}
 Note that $V_H$ is even, but can be made heavy (with mass of order $\Lambda$) by taking $g_C$ large. The Standard Model $SU(2)_L$ is the diagonal subgroup of the $3$ $SU(2)$'s and its gauge coupling is given 
\begin{equation}
\frac{1}{g_{\text{SM}}^2} = \frac{1}{g_A^2} + \frac{1}{g_B^2} + \frac{1}{g_C^2} = \frac{2}{g^2} + \frac{1}{g_C^2} \;.
\end{equation}
In the limit where $g_C \gg 1$, we have :
\begin{equation}
g_{\text{SM}} = \frac{g}{\sqrt{2}} \; \;.
\end{equation}
To find the action of T-parity on the field $\Sigma$ it is useful to think of it  as a composite of strongly interacting fermions:
\begin{equation}
\Sigma \sim  \Psi \Psi^T
\end{equation}
where $\Psi$ form a 6 of the $SU(6)$ flavor symmetry. The quantum numbers of the different components of $\Psi$ are as follow:
\begin{center}
    \begin{tabular}{ | l | | l | | l | | l | | l | | l | | l |  | p{4cm} | }
    \hline
  Field  &$   SU(2)_A $  & $   SU(2)_B $  & $ U(1)_Y $     \\ \hline
 $   \psi_{2_A} $  &  2 &   1 &  $  0$    \\ \hline  
   $   \psi_0 $ & 1  &  1&  $ \frac{1}{2} $  \\ \hline
     $  \psi_0' $ &1 &  1 &  $-\frac{1}{2} $  \\  \hline
          $  \psi_{2_B} $ &1 &  $2^*$ &  $0$   \\  \hline
\hline
  \end{tabular}
  \end{center}
  with two $SU(2)$ doublets and two singlets, for a total of six fields. The T-parity action is then:
  \begin{equation}
  \psi_{2_A} \rightarrow  \epsilon \psi_{2_B} \; \; \; \;  \psi_0 \rightarrow \psi_0 \; \;  \; \; \psi_0'  \rightarrow \psi_0'  \; \; \; \; \psi_{2_B} \rightarrow - \epsilon \psi_{2_A} \; \;.
  \end{equation}
  This leads to the following transformation for the non-linear sigma model field at low energy:
  \begin{equation}
  \Sigma \rightarrow T \Sigma T^{T} 
\label{eq:tactionS},
\end{equation}
where $T$ is an element of $Sp(6)$ and is given by:
\begin{equation}
 T =\begin{pmatrix}    0  &  0& \epsilon\\ 
0 &  \eins   & 0  \\ 
  - \epsilon &0 &   0  \\ 
 \end{pmatrix} .
\end{equation}
This in turn leads to the following action on the components of the $\Pi$ field:
\begin{align}
\nonumber
& \eta \rightarrow \eta,  \\ \nonumber
& \tilde{\phi} \rightarrow - \tilde{\phi},  \\ \nonumber
& \chi \rightarrow  - \chi^{*} , \\ \nonumber
& h_A \rightarrow - \epsilon h_B^* .
\end{align}
T-even and T-odd Higgs doublets can then be defined:
\begin{eqnarray}
\label{eq:evenoddhiggses}
H &=& \frac{h_B + \epsilon h_A^*}{\sqrt{2}} \; \;, \\ \nonumber
\tilde{H} &=& \frac{h_B - \epsilon h_A^*}{\sqrt{2}} \; \; ,
\end{eqnarray}
where here and in the rest of the paper, the fields with a {\it tilde} are odd under T-parity. The T-parity action on the Higgs fields will lead to an inert doublet Higgs model \cite{Barbieri:2006dq} as will discuss in more details later. The $SU(2)$ triplets $\tilde{\phi}$, $\phi_A$ and $\phi_B$ form two T-odd and one T-even combination. Two of them, one even and one odd, are eaten by the massive gauge boson $V^\mu_H$ and $\tilde{V}^\mu$, while the remaining odd combination:
\begin{equation}
\tilde{\Phi} = \frac{\phi_A - \phi_B - \tilde{\phi}}{\sqrt{3}}
\end{equation}
will be part of the low energy spectrum.

\subsection{Top sector}
\label{sec:top}
As  in most little Higgs models, the third generation of quarks is treated differently than the other fermions as only the top Yukawa is large enough to introduce a fine-tuning problem in a model with a UV cutoff at $10$ TeV. As such it is the only one that needs to be introduced in a way that respect the collective symmetry breaking mechanism.  There is some freedom on how to do that as one can choose to keep a very minimal particle content that respect only a small subset of the global symmetries, or opt for a more symmetric coupling between the fermions and the non-linear sigma model field. In the most symmetric case, $Q_3 $ and $Q_3^c$, two sixplets of fermions are introduced and coupled to $\Sigma$ in an $SU(6)$ invariant way.  This create many massive vector-like fermions and massless combinations can be recovered by adding $q_C$, a $SU(2)_C$ doublet and a singlet ${u'}^c$. The Lagrangian is given by:
  \begin{equation}
\mathcal{L}_{\text{top}}^{SU(6)} = f \left[ \lambda_1  {Q_3}^{T}  \Sigma^{\dagger}  Q_3^c+   \lambda_2 \left( -q_C^T  K_1^{T} \epsilon q_A^c  +q_C^T  K_2^{T}  q_B^c \right) + \lambda_3  u  \tilde u^c \right] + \text{h.c.} \;,
\end{equation}
where  $q_A^c, q_B^c$ and $q_C$ form $SU(2)$ doublets and $Q_3$ and $Q_3^c$ are sixplet of the global $SU(6)$:
\begin{align}  
  Q_3=\begin{pmatrix}   q_A  \\ 
 u  \\ 
  d  \\ 
   q_ B  \\ 
 \end{pmatrix} ,
  & \ \
     Q_3^c =\begin{pmatrix}   q^c_A  \\ 
 d^c  \\ 
  u^c  \\ 
   q^c_B  \\ 
 \end{pmatrix} .
  \end{align}
Under T-parity we have:
\begin{equation}
Q_3 \rightarrow T Q_3 \; \; \; \; Q_3^c \rightarrow T Q_3^c \; \; \; \; \tilde{u}^c \rightarrow \tilde{u}^c \;\;\;\; q_C \rightarrow q_C .
\end{equation}
Replacing $\Sigma$, $K_A$ and $K_B$ by their {\it vev} (but setting the Higgses {\it vev} to 0), we get:
\begin{equation}
\mathcal{L}_{\text{top}}^{SU(6)}  = \lambda_1 f \left( q q^c + \tilde{q} \tilde{q}^c  + d d^c \right)+  \frac{\lambda_2}{\sqrt{2}} f q_C q^c  +f u \left(\lambda_3 \tilde{u}^c - \lambda_1 u^c \right),
\end{equation}
where:
\begin{eqnarray}
q = \frac{q_A + \epsilon q_B}{\sqrt{2}} &&  q^c = \frac{q_B^c - \epsilon q_A^c}{\sqrt{2}}  \\ \nonumber
\tilde{q}  = \frac{q_A - \epsilon q_B}{\sqrt{2}} & & \tilde{q}^c = \frac{-q_B^c - \epsilon q_A^c}{\sqrt{2}}
\end{eqnarray}
So we have two massive ( with mass of order $f$) vector-like doublets and two massive vector-like T-even singlets. The remaining T-even  doublet and singlet are massless at this point. They are to first approximation the Standard Model top quark doublet and singlet, and get a mass after electroweak symmetry breaking. The masses of the heavy fields are:
\begin{eqnarray}
\label{eq:topmasses}
m_{Q} = \sqrt{\lambda_1^2 + \lambda_2^2} f ,& & m_{\tilde{Q}}= \lambda_1 f, \\ \nonumber
m_{U} =  \sqrt{\lambda_1^2 + \lambda_3^2} f ,& & m_D = \lambda_1 f \;,
\end{eqnarray}
and the mass eigenstates by:
\begin{eqnarray}
q_\text{SM} &=& \frac{\lambda_2 q - \lambda_1 q_C}{\sqrt{\lambda_1^2 + \lambda_2^2}} = \sin \left(\theta_D \right) q - \cos \left(\theta_D\right) q_C,\\ \nonumber
u_\text{SM}^c &=& \frac{ \lambda_1 \tilde{u}^c + \lambda_3 u^c}{\sqrt{\lambda_1^2 + \lambda_3^2}}= \cos \left(\theta_S\right) {u'}^c + \sin \left( \theta_S\right) u^c, \\ \nonumber
Q &=& \cos \left( \theta_D\right)  q + \sin \left( \theta_D\right) q_C, \\ \nonumber
U^c &=& \sin \left(\theta_S\right) {u'}^c - \cos \left( \theta_S \right) u^c \;.
\end{eqnarray} 
From these mass eigenstates we can read off the top Yukawa coupling which is given by:
\begin{equation}
\label{eq:topyukawa}
\lambda_t = \frac{\sqrt{2} \lambda_1 \lambda_2 \lambda_3}{\sqrt{\lambda_1^2 + \lambda_2^2} \sqrt{\lambda_1^2+\lambda_3^2}} = \frac{ \sin 2 \theta_D\sin \theta_S m_{Q}}{\sqrt{2}f} \; \;.
\end{equation} 
Note that this implies a lower bound on $m_Q$:
\begin{equation}
m_Q > \sqrt{2} \lambda_t f \; \; .
\end{equation}

The main issue with this top sector is that it contains the T-even $d^c$ state which can lead to large contribution to $Z \rightarrow b \bar{b}$ when exchanged at tree-level\cite{Gregoire:2003kr}. There are various possibilities to remedy this problem. For example, $d^c$ can be removed from $Q_3^c$. The top sector Lagrangian then becomes:
\begin{equation}
\mathcal{L}_{\text{top}}^{SU(6)_L \times SU(5)_R} = \lambda_1 f  \begin{pmatrix}q_A & u & d & q_B \end{pmatrix} \Sigma^\dagger \begin{pmatrix} q_A^c \\ 0 \\ u^c \\q_B^c \end{pmatrix}+ f \frac{\lambda_2}{\sqrt{2}} \left( -q_C^T  K_A^{T} \epsilon q_A^c  +q_C^T  K_B^{T}  q_B^c \right) + \lambda_3 f u \tilde{u}^c + \lambda_4 f d d^c
\end{equation}
This Lagrangian contains an additional free parameter $\lambda_4$, but the dangerous coupling of $d^c$ to the Higgs is removed. The masses of the various states are still given by \eqref{eq:topmasses} except for $m_{D}$ which becomes: 
\begin{equation}
 m_D = \lambda_4 f \; \;.
\end{equation}
The top Yukawa remains is also still given by equation \eqref{eq:topyukawa}.

Another possibility is to remove the down-like states completely as in the $SU(5)_{MLS}$ top sector of \cite{Gregoire:2003kr}. Being less symmetric, the $SU(5)_{MLS}$ and $SU(6)_L \times SU(5)_R$ top sectors lead to larger deviation to the $T$ parameter as we will see in section \ref{sec:ewpm}. Also, the radiative corrections to the Higgs potential coming from the $SU(6)$ top sector is one-loop finite and calculable. In the $SU(6)_L \times SU(5)_R$ top sector, only the Peccei-Quinn violating $b^2$ term is finite and calculable, while in the $SU(5)_{MLS}$ top sector, both the masses and $b^2$ term are log-divergent at one-loop and therefore not calculable. 
\subsection{ Light Fermions}  
\label{sec:fermions}
While the top quark Yukawa coupling needs to be introduced in a way that respects collective symmetry breaking, this is not true for the other fermions. They can arise from two doublets $  q_A \in 2_{SU(2)_A} $   and $ q_B \in 2_{SU(2)_B} $ which can be embedded in a  $SU(6) $  incomplete multiplet:
  \begin{align}  
  Q_{1,2} =\begin{pmatrix}     q_A  \\  
0 \\ 
  0  \\ 
   q_B  \\ 
 \end{pmatrix} ;
   \end{align}
on which the T parity acts as:
 \begin{align}
Q_{1,2} \rightarrow T Q_{1,2} ,
\end{align} 
which means
  $  q_A \rightarrow \epsilon q_B.$ 
  The T parity eigenstates are the odd linear combination:   $ \tilde{q}= \frac{ q_A - \epsilon q_B}{\sqrt{2}} $ and
   the even one: $q_{\text{SM}} = \frac{ q_A+ \epsilon q_B}{\sqrt{2}} $ which is identified with the SM doublet. 
\newline  
 In order to give mass to the odd doublet we introduce a  $ SU(2)_C $ doublet  $ \tilde{q}_C^c  $  odd under T parity  $ \tilde{q}_C^c \rightarrow  - \tilde{q}_C^c$ which can marry the $\tilde{q}$ with the help of the link fields:    
 \begin{equation}
 \frac{k}{\sqrt{2}} f_k [ q_A^T   K_A^*  + q_B ^T \epsilon K_B^* ] \tilde{q}_C^c ;
\end{equation}
which gives mass to the odd Dirac pair:
\begin{align}
 m_{\tilde{ q}}=  k f_k , \end{align}
leaving the T even linear combination massless.

The Yukawa interactions for the light fermions can be written down by introducing the SM singlets  simply   as T-even fields embed  in incomplete $ SU(6) $ multiplets :
  \begin{align}  
 \U_{1,2}^c =\begin{pmatrix}    0  \\ 
u^c \\ 
  0\\ 
   0  \\ 
 \end{pmatrix} ,
 &  \  
 \D_{1,2}^c =\begin{pmatrix}    0  \\ 
0 \\ 
 d^c \\ 
   0  \\ 
 \end{pmatrix} .
   \end{align}
 The up and down type Yukawa coupling can then be written as:
   \begin{equation}
  k_u  f Q_{1,2}^{T}  \Sigma^{\dagger} \U_{1,2}^c + k_d f Q_{1,2}^T \Sigma^\dagger \D_{1,2}^c
  \end{equation}
  and the full Lagrangian for the light standard model fermions becomes:
  \begin{equation}
  \label{eq:Llight}
 \mathcal{L}_{\text{light}} =  k_u  f Q_{1,2}^{T}  \Sigma^{\dagger}  \U_{1,2}^c + k_d  f Q_{1,2}^{T}  \Sigma^{\dagger} \D^ c+\frac{k}{\sqrt{2}} f_k [ q_A^T   K_A^*  + q_B ^T \epsilon K_B^* ] \tilde{q}_C^c +h.c..
\end{equation}

We note that the fermion content presented here contains $SU(2)_i^2 U(1)_Y$ anomalies (with $i=A,B,C$). It is, however, straightforward to add heavy fields, with mass of order $\sim 10$ TeV, to cancel them.

\section{Radiative corrections}
\label{sec:radcorr}
The gauge and Yukawa interactions introduced in the previous section will radiatively induce additional contributions to the Lagrangian of the theory. The form and size of the new terms depend on the way the interactions involved in the radiative corrections break the various global symmetries of the theory. For example, turning on only one gauge coupling, or only one top Yukawa coupling preserve enough of global symmetry so that the Higgs remains an exact Goldstone boson and contributions proportional to, for example, $g^2$ cannot include a Higgs mass term. At one loop, the radiative corrections can be quadratically divergent, log-divergent or finite. In the case of quadratic or log divergence, this indicates that the correction is sensitive in the UV and cannot be fully computed in the low energy theory. The corresponding operator should be included in the Lagrangian with an unknown coefficient. However the size of the one-loop quadratic or log-divergent diagram constitute an natural estimate for the coefficient of this operator\footnote{ more precisely, a spurion analysis has to be performed, to make sure that, for example that the size of the one-loop log-divergent operator is not parametrically smaller than the size of an operator generated via a two-loops quadratically divergent operator.}. In the case where terms in the effective Lagrangian only appear via finite one-loop diagram, this means that they are dominated by IR physics and are therefore computable. There might be divergent contributions to the same terms at higher loop order, but these will be subdominant.  We now examine the size of the principal radiative corrections.
\subsection{One-loop quadratic divergence}
At one loop, gauge interactions will generate terms in the potential that are quadratically divergent:\begin{equation}
V_{g^2}=  \frac{g^2}{16 \pi^2} f^2 \Lambda^2 \left[ c_1 \text{Tr} \left( \Sigma Q_A^a \Sigma^\dagger Q_A^a \right) + c_2 \text{Tr} \left( \Sigma Q_B^a \Sigma^\dagger Q_B^a \right) \right] \;.
\end{equation}
 Computation of the Coleman-Weinberg potential \cite{Coleman:1973jx} suggests that $c_1$ and $c_2$ are of order $1$ and $T$-parity guarantees that $c_1=c_2 =c$. The Coleman-Weinberg potential computation actually yield a negative value for $c$ which if correct would signal that we are in the wrong vacuum \cite{Piai:2004yb}. However,  the coefficients of those operators are not calculable in the low energy theory and their value (and their sign) depend on the UV completion.  We will assume that they are positive and that we are in the correct vacuum, which according to the work of \cite{Piai:2004yb} would imply that the UV completion, unlike QCD,  does not have a 'vector limit'  in the sense of Georgi\cite{Georgi:1989xy}. We also note that the estimate $c \sim 1$  rely on taking $\Lambda \sim 4 \pi f$, which might be an overestimation \cite{Chang:2003vs}. But once again, the value of $c$ is a UV dependent quantity, and we will work under the assumption that it is indeed of order $1$ such that the two operators above yield a mass of order $f \sim 1 \text{TeV}$ for the singlet $\chi$ and an order one quartic couplings for the two Higgs doublets:
\begin{equation}
\label{eq:opquartic}
V_{g^2} = c f^2 \left|\chi - \frac{ h_A^T \epsilon h_B }{f^2}  \right|^2 + c f^2  \left| \chi  + \frac{  h_A^T \epsilon h_B }{f^2}  \right|^2 = c \left[ f^2 \left| \chi \right|^2 + \left| h_A^T \epsilon h_B \right|^2 \right] \; \; \;.
\end{equation}

 The other operator to receive a 1-loop quadratically divergent contribution comes from the hypercharge sector:
  \begin{equation}
  \label{eq:opdivY}
 V_{g'^2}= c_Y \frac{3 g'^2}{16 \pi^2} f^2 \Lambda^2 \text{Tr} \left( \Sigma Y \Sigma^\dagger Y \right) = c_Y f^2 \Lambda^2 \frac{3 g'^2}{16 \pi^2} \left[  \left( \frac{\left| h_A \right|^2 + \left| h_B \right|^2}{f^2} \right) + \frac{2}{3} \left( \frac{\left| h_A\right |^2 + \left| h_B \right|^2}{f^2}  \right)^2 \right] + \cdots
 \end{equation}
 with $c_Y \sim 1 $.  Since a collective breaking mechanism was not implemented in this sector, it contains a mass term for the Higgses. Being quadratically divergent, the size of this operator is not calculable, nevertheless, taking $c_Y \sim 1$ will lead to an appropriately small Higgs mass, owing to the smallness of the $U(1)_Y$ gauge coupling. In fact, with a Higgs mass of $120$ GeV and a cutoff of $\Lambda=10 $TeV, the fine-tuning due to the hypercharge one loop quadratic divergence is about 1 part in 4 (see \cite{Barbieri:2006dq}).
 
 \subsection{One-loop log divergence}
A few operators will be generated with coefficient proportional to $g_A^2 g_B^2 = g^4$. Being proportional to two weak coupling constants they have coefficients parametrically smaller than (\ref{eq:opdivY}), but they provide one of the dominant contribution to the Higgs mass. For example, the one loop log-divergent contribution to Higgs potential is given by:
\begin{equation}
\label{eq:oploggauge}
V_{g^4}(h_A,h_B)= \frac{9 g_A^2 g_B^2}{16 \pi^2} \left[ f^2 \left(\left| h_A \right|^2 + \left| h_B \right|^2 \right) - \left( \frac{5}{6} \left(\left|h_A \right|^2 + \left|h_B\right|^2 \right)^2 - \frac{2}{3} \left|h_A^T \epsilon  h_B \right|^2 \right)  \right] \log \frac{\Lambda^2}{M_{\tilde{V}}^2} \; \; .
 \end{equation}
 Note that the Higgs quartic couplings generated in \eqref{eq:opquartic} , \eqref{eq:opdivY} and \eqref{eq:oploggauge}  all respect an extra symmetry that guarantees that the charged Higgs and CP-even neutral Higgs will have the same mass. The $SU(2)'s$ gauge couplings leave unbroken a custodial $SU(2)_{\text{custodial}}$ under which the $2 \times 2$ matrix $\mathcal{H}$ with $h_A$ and $h_B$ as columns transforms as: 
  \begin{equation}
 \mathcal{H} \rightarrow U \mathcal{H} U^\dagger \; \; \; \text{with } \; \; \; U \in SU(2)_{\text{custodial}} \; \;.
  \end{equation}
 Together with a Pecci-Quinn symmetry, this guarantees that the operators generated only through $SU(2)_1$ and $SU(2)_2$ gauge couplings will produce quartic couplings of the form:
 \begin{equation}
 c_1 \left( \left|h_A \right|^2 + \left| h_B \right|^2 \right)^2 + c_3 \left| h_A^T \epsilon  h_B \right|^2
 \end{equation}
 On the other hand, the hypercharge gauge coupling leaves  an $Sp(4)$ symmetry invariant (acting on indices $1,2,5$ and $6$ of the $\Sigma$ field) which forces the Higgs quartic interaction to be of the form 
 \begin{equation}
 \left( \left|h_A \right|^2 + \left| h_B \right|^2 \right)^2 \; \;.
 \end{equation}
The leading quartic interaction that will induce a difference in mass difference between the charged and CP-even, T-odd Higgses come from operators generated with coefficient proportional to $g^2 g'^2$. At one loop there is a log divergent contribution given by:
\begin{equation}
\label{eq:oplogGGY}
V_{g^2 g'^2} (h_A,h_B) = -\frac{3 g^2 g'^2}{16 \pi^2} \left[ \left( \left|h_A \right|^2 - \left| h_B \right|^2 \right)^2+ 4 \left| h_A^T \epsilon h_B \right|^2 \right] \log \frac{\Lambda^2}{m^2_{\tilde{V}}}\; \; .
\end{equation}
 
 \subsection{One-loop finite correction}
 As mentioned above, the one-loop log divergent and quadratically divergent contributions to the potential are not calculable. They are sensitive to the UV completion and they correspond to operators that have to be included in the Lagrangian with unknown coefficients. For example, in addition to 1-loop log-divergent contribution proportional to $g_A^2 g_B^2$, there might also be two loop quadratically divergent contributions with the same size.
  
  There are also terms which get their main contribution from one loop finite term and are therefore calculable. The odd triplet $\tilde{\Phi}$ for example gets a calculable mass given by:
 \begin{equation}
  M^2_{\tilde{\Phi}} = \frac{6}{( 4 \pi)^2} \frac{f^2 f_k^2 g_A^2 g_B^2 g_C^2 }{ ( M^2_{V_H}-M^2_{\tilde{V}})} 
  \log{\frac{ M_{V_H}}{ M_{\tilde{V}}}} .
 \end{equation}
 Note however that as $g_C$ approaches $4 \pi$ and $V_H$ becomes heavy, pertubativity in $g_C$ is lost, and while we expect the size of this mass to remain of the same order, it will not be calculable anymore.
 
  The $SU(6)$ top sector has extra spurious symmetries and its dominant contribution to the Higgs potential is a one loop finite piece. This sector also breaks the PQ symmetry, so the quadratic term will also contain a '$b^2$ term': $h_A^T \epsilon h_B$.  It is in fact the dominant contribution to the $b^2$ term which is therefore calculable:
  \begin{equation}
 V_{\text{top}}(h_A,h_B) = -\frac{3}{8 \pi^2} \lambda_t^2 \frac{m_{Q}^2 m_U^2}{m_{Q}^2-m_U^2} \log\left(\frac{m_{Q}^2}{m_U^2} \right) \left[\left|h_A\right|^2 + \left|h_B \right|^2 - \left(h_A^T \epsilon h_B + \text{h.c.} \right) \right]
   \end{equation}
   
\begin{figure}
\begin{center}
\psfrag{labely}{\footnotesize $b$(GeV)}
\psfrag{labelx}{\footnotesize $m_{\tilde{Q}}$(GeV)}
\psfrag{label11}{\footnotesize $1.1$ TeV}
\psfrag{label13}{\footnotesize $1.3$ TeV}
\psfrag{label15}{\footnotesize $1.5$ TeV}
\includegraphics[width=7.5cm]{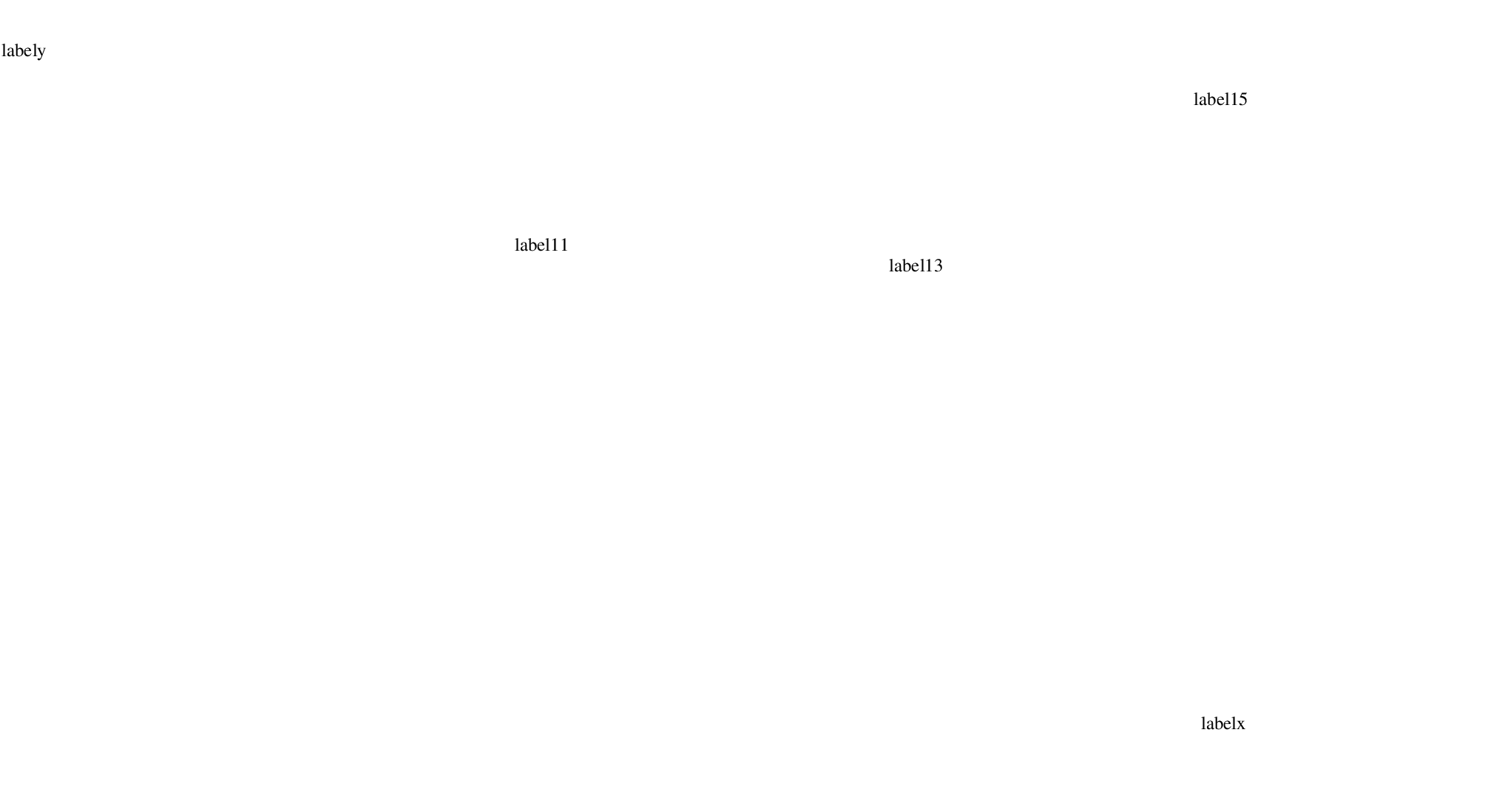}
\includegraphics[width=7.5cm]{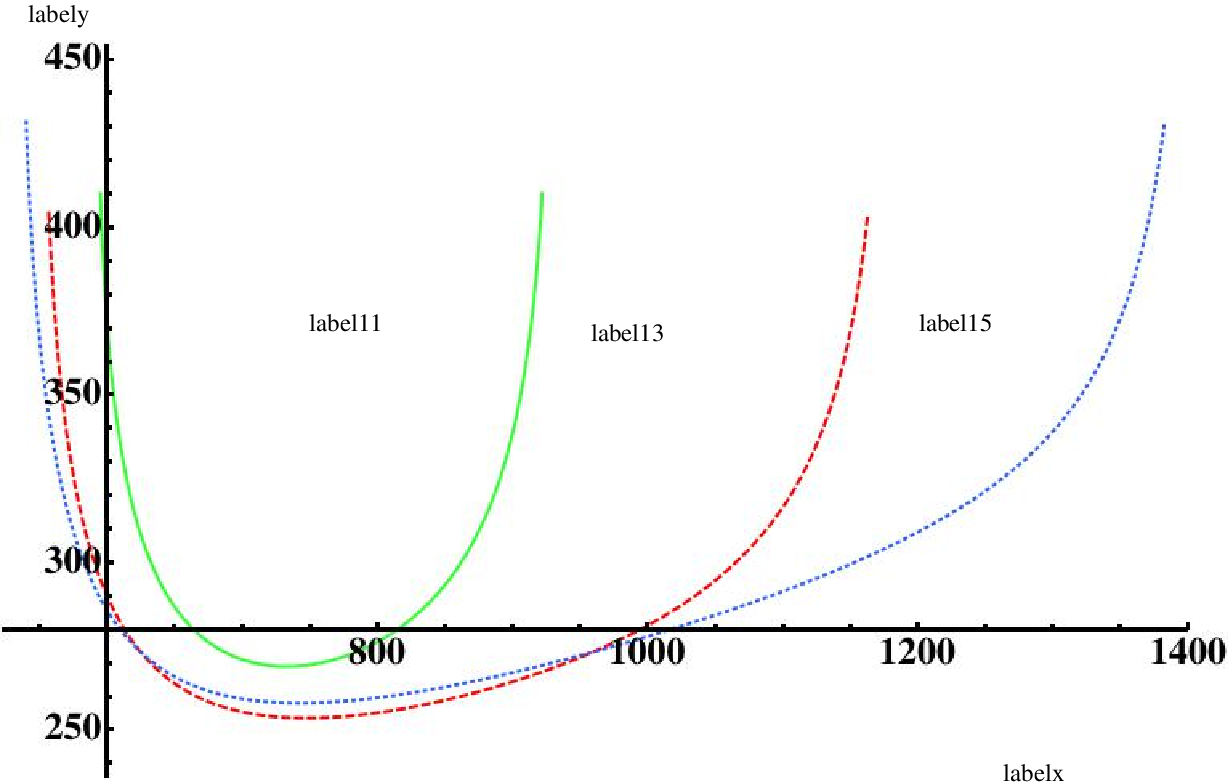}
\caption{b as a function of $m_{\tilde{Q}}$, for various choices of $m_{U}$ with and $f=700$ GeV. The plot on the left is for the $SU(6)_L \times SU(5)_R$ top sector with $m_{D} = 1$ TeV and the plot on the right is for the $SU(6)$ top sector.}
\label{fig:b}
\end{center}
\end{figure}

   For the $SU(6)_L \times SU(5)_R$ top sector, with $d^c$ removed from $Q_3^c$, the 1-loop contribution to $\left|h_A \right|^2 + \left|h_B \right|^2$ is log-divergent, but the coefficient of the  $b^2$ term remains finite and calculable. It is given by:
   \begin{equation}
   b^2= -\frac{3}{4 \pi^2} \left[\lambda_1^2 \left( \frac{m_Q^2 m_D^2}{m_Q^2-m_D^2} \log \frac{m_D^2}{m_{Q}^2} + \frac{m_{\tilde{Q}}^2 m_D^2}{m_{\tilde{Q}}^2 - m_D^2} \log \frac{m_{\tilde{Q}}^2}{m_D^2}  \right) + \lambda_t^2 \frac{ m_{Q}^2 m_U^2}{2(m_{Q}^2 - m_U^2)} \log \frac{m_U^2}{m_{Q}^2} \right] \;.
   \end{equation}
 The value of $b^2$ as a function of $m_{\tilde{Q}}$ is shown in figure \ref{fig:b} for $f=700$ GeV and various choices of quark masses and for both top sectors. 

We note that the singlet $\eta$ does not get any mass from radiative corrections.  It can be given a small mass explicitly \cite{Gregoire:2003kr}.
 \section{Electroweak symmetry breaking}
\label{sec:ewsb}
 We now examine the Higgs potential in more details. Because of the T-parity, the Higgs potential will be of the inert doublet type \cite{Barbieri:2006dq, Ma:2006km}. In the $h_A$ and $h_B$ basis, the most general potential that respect T-parity is given by:
 \begin{multline}
 \label{eq:inertAB}
 V(h_A,h_B) = m^2 \left(\left|h_A\right|^2 + \left|h_B \right|^2 \right) + b^2 \left(h_A^T \epsilon h_B + \text{h.c.}\right) \\
 +c_1 \left( \left|h_A \right|^4 + \left|h_B \right|^4 \right) +c_2 \left|h_A \right|^2 \left|h_B \right|^2 +c_3 \left|h_A^T \epsilon h_B \right|^2 +\\
  \frac{c_4}{2} \left[ \left(h_A h_B \right)^2 + \text{h.c.} \right] + \frac{c_5}{2} \left[ \left(h_A h_B\right) \left(\left|h_A \right|^2 + \left|h_B\right|^2 \right) + \text{h.c} \right]
 \end{multline}
  In our model, we expect $c_3$ to be the dominant coupling as it receives an order $1$ contribution from the quadratic divergence of equation (\ref{eq:opquartic}). The coefficients $c_1$ and $c_2$  also receive contribution from a quadratic divergence from the hypercharge sector (see equation (\ref{eq:opdivY})) but due to the smallness of $g'$ we expect them to be somewhat smaller than $c_3$. Contributions to $c_2 - 2 c_1$ are of order 1-loop and come from log-divergent contributions such as the one given in equation (\ref{eq:oplogGGY}) or possibly from the top sector:
\begin{equation}
c_2 - 2 c_1  \equiv \delta \sim \frac{1}{16 \pi^2} \; \;.
\end{equation}
The Peccei-Quinn violating quartic $c_4$ and $c_5$ are also only generated at one loop.

 To minimize the potential and find the mass eigenstates, it is useful to write it in term of the  T-parity eigenstate  $H, \tilde{H}$ (see \eqref{eq:evenoddhiggses}):
 \begin{multline}
 V(H,\tilde{H}) = \mu_+^2 \left|H \right|^2 + \mu_-^2 \left|\tilde{H}\right|^2 \\
  + \lambda_1 \left|H \right|^4 + \lambda_2 \left|\tilde{H} \right|^4 + \lambda_3 \left|H \right|^2 \left|\tilde{H}\right|^2 + \lambda_4 \left| H^\dagger \tilde{H} \right|^2 + \frac{\lambda_5}{2} \left[ \left(H^\dagger \tilde{H} \right)^2 + \text{h.c.} \right]
 \end{multline}
 
 with
 \begin{eqnarray}
 \mu^2_+ &=& m^2 - b^2 \; \; ,\\ \nonumber
 \mu^2_- &=& m^2 + b^2\; \; ,\\ \nonumber
 \lambda_1 &=& \frac{c_1}{2} +\frac{c_2}{4} + \frac{c_3}{4}+\frac{c_4}{4}+\frac{c_5}{2}\;\; , \\ \nonumber
  \lambda_ 2 &=&\frac{c_1}{2} +\frac{c_2}{4} + \frac{c_3}{4}+\frac{c_4}{4}-\frac{c_5}{2} \;\; ,\\ \nonumber
 \lambda_3 &=&c_1 +\frac{c_2}{2} -\frac{c_3}{2}-\frac{c_4}{2}\; \; ,\\ \nonumber
 \lambda_4 &=&c_1 -\frac{c_2}{2} +\frac{c_3}{2}-\frac{c_4}{2}\; \; , \\ \nonumber
 \lambda_5 &=&c_1 -\frac{c_2}{2}-\frac{c_3}{2} + \frac{c_4}{2}\; \; .\\ \nonumber
 \end{eqnarray}
 This potential has an extremum where only $H$ gets a vev and T-parity is unbroken:
 \begin{equation}
 \left< H \right> = \begin{pmatrix} 0 \\ \frac{v}{\sqrt{2}} \end{pmatrix} \; \; \; \; \left< \tilde{H} \right> = \begin{pmatrix} 0 \\ 0 \end{pmatrix} \; \; .
 \end{equation}
 At this point $H$ is identical to a Standard Model Higgs doublet, with 3 of its components eaten by the $W$ and $Z$ boson and the remaining scalar $h$ being the usual neutral Higgs boson. The T-odd doublet $\tilde{H}$ contains a charged scalar $\tilde{h}^+$, a neutral CP odd scalar $\tilde{A}$ and a neutral CP even scalar $\tilde{S}$. The masses for these fields are given by \cite{Dolle:2009fn}:
 \begin{eqnarray}
 \label{eq:higgsmass}
 m_h^2 &=& -2 \mu_+^2 = 2 \lambda_1 v^2 = 2(b^2-m^2),\\
 m_{\tilde{h}^+}^2 &=& \mu_-^2 + \lambda_3 \frac{v^2}{2},\\
 m_{\tilde{S}}^2 &=& \mu_-^2 + \left(\lambda_3 + \lambda_4 + \lambda_5  \right) \frac{v^2}{2}, \\
 m_{\tilde{A}}^2 &=& \mu_-^2 + \left(\lambda_3 + \lambda_4 - \lambda_5  \right) \frac{v^2}{2} .
 \end{eqnarray}
As long as the parameters are such that these masses are positive, this vacuum will be a minimum of the potential. With only the dominant quartic coupling $c_3$ non-zero we have: $\lambda_1 = \lambda_2 = -\lambda_3/2 = \lambda_4/2= -\lambda_5/2$ which implies the following relationships among the masses:
\begin{eqnarray}
\label{eq:msvsmhp}
m^2_{\tilde{h}^+} &=& m^2_S  \; \; , \\
\label{eq:mavsmsmh}
m^2_{\tilde{A}}  &=& m_{\tilde{S}}^2 + m_h^2 \; \; .
\end{eqnarray}
In term of the parameters in \eqref{eq:inertAB} we have $m_{\tilde{S}}^2 = m_{\tilde{h}^+}^2 = 2m^2$ and $m_{\tilde{A}}^2 = 2 b^2$, where the second relation always hold if the Peccei-Quinn violating couplings $c_4$ and $c_5$ are $0$. Since $m_h >0$ requires $b>m$, we find that $\tilde{A}$, the T and CP odd scalar is the heaviest among the Higgs states. The presence of a non-zero $c_1$ and $c_2$ will cause $m_{\tilde{h}^+}^2$ and $m_{\tilde{S}}^2$ to deviate from $2 m^2$ while a deviation from the relation $c_2 = 2 c_1$ will lead to $m_{\tilde{h}^+} \neq m_{\tilde{S}}$:
\begin{equation}
\label{eq:mhpmsdelta}
m_{\tilde{h}^+}^2 = m_{\tilde{S}}^2 + \frac{\delta}{2} v^2 \; \; .
\end{equation}
Since we want $\tilde{S}$ to be the dark matter candidate, we require $m_{\tilde{h}^+} > m_S$ which in turnÊ require $2 c_1 > c_2$.  Finally we also note that we expect the Higgs mass $m_h$ to be of the same order of $b$, unless there is some fine tuning between $m$ and $b$, as seen from equation \eqref{eq:higgsmass}.

\section{Electroweak  Precision Observables}
In this section we examine contributions to electroweak precision observables coming from states that are the in the low energy theory, below $\Lambda$. There are also uncalculable contributions from the UV completion in the form of higher-dimensional operators suppressed by a scale $\Lambda$. If this scale is $\sim 4 \pi f \sim 10$ TeV, those contributions are not significant \cite{Barbieri:2000gf}. If the scale suppressing the higher dimensional operator is smaller, as the unitarity studies suggest \cite{Chang:2003vs}, the effects of those operator should be included in a global analysis. 
\label{sec:ewpm}
\subsection{Higgs sector contributions}
Because of the T-parity, there are no tree level contributions to electroweak precision observables. There are however 1-loop contributions to the $S$ and $T$ parameters \cite{Peskin:1991sw}. The largest contributions will come from 'light' fields with mass of order $f/(4 \pi) $ while contributions from loops of 'heavy' field, with mass of order $f$ are parametrically suppressed, but can give numerically large contributions in the case of the top sector . The 'light' fields that could contribute are the odd triplet $\tilde{\Phi}$ and the odd Higgses. The first gives a small contribution because it has a small coupling to the SM Higgs. In the case of loops of the odd Higgses, $\tilde{S}$, $\tilde{h}^+$ and $\tilde{A}$, the contribution to  $T$ is small due to the approximate custodial symmetry of the Higgs potential (which leads to the degeneracy in $\tilde{S}$ and $\tilde{h}^+$). 

\begin{figure}
\begin{center}
\psfrag{labely}{$S$}
\psfrag{labelx}{$m_{\tilde{A}}$(GeV)}
\psfrag{label120}{$120$ GeV}
\psfrag{label200}{$200$ GeV}
\psfrag{label300}{$300$ GeV}
\includegraphics[width=7.5cm]{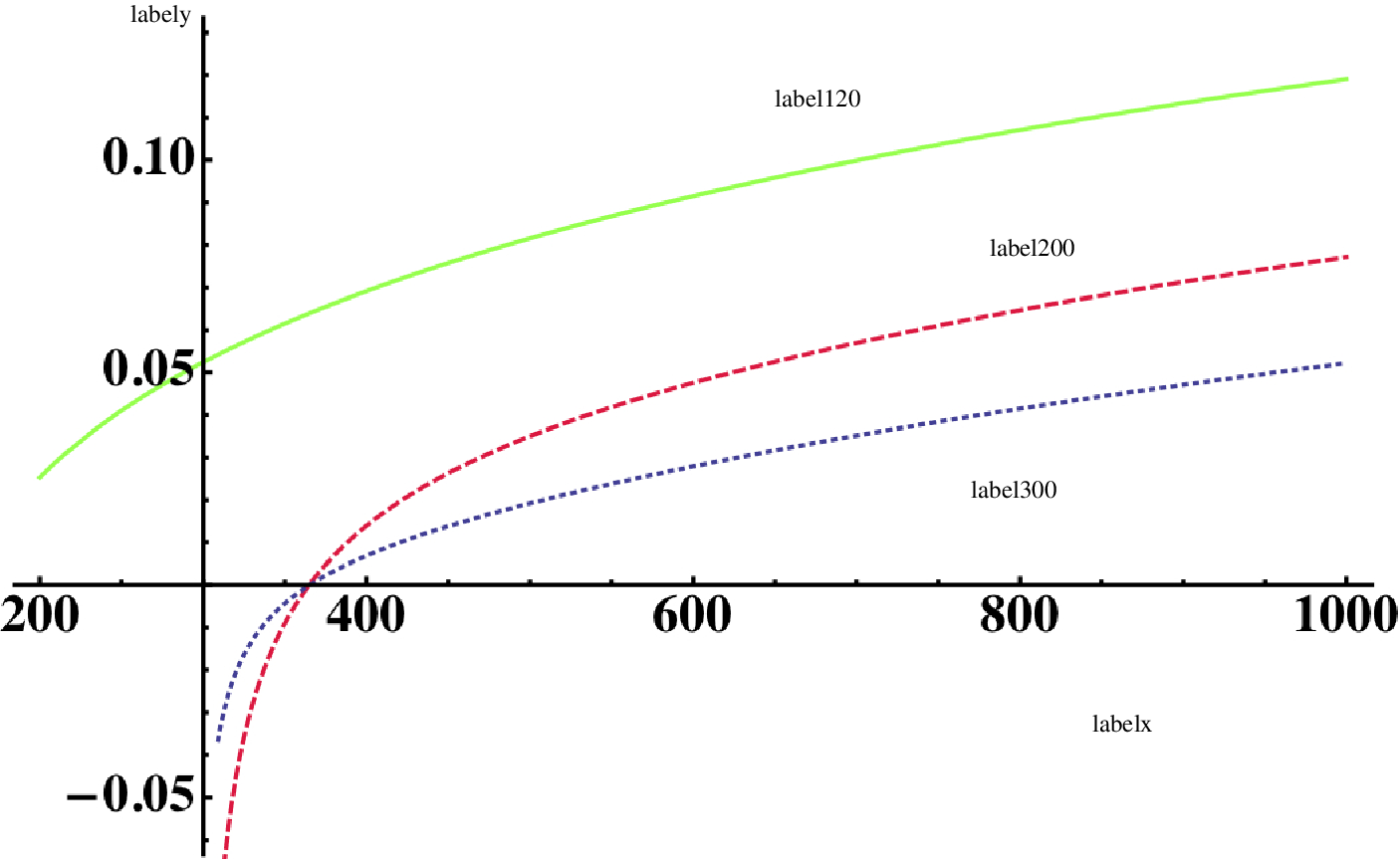}
\psfrag{labelx}{$m_{\tilde{S}}$(GeV)}
\includegraphics[width=7.5cm]{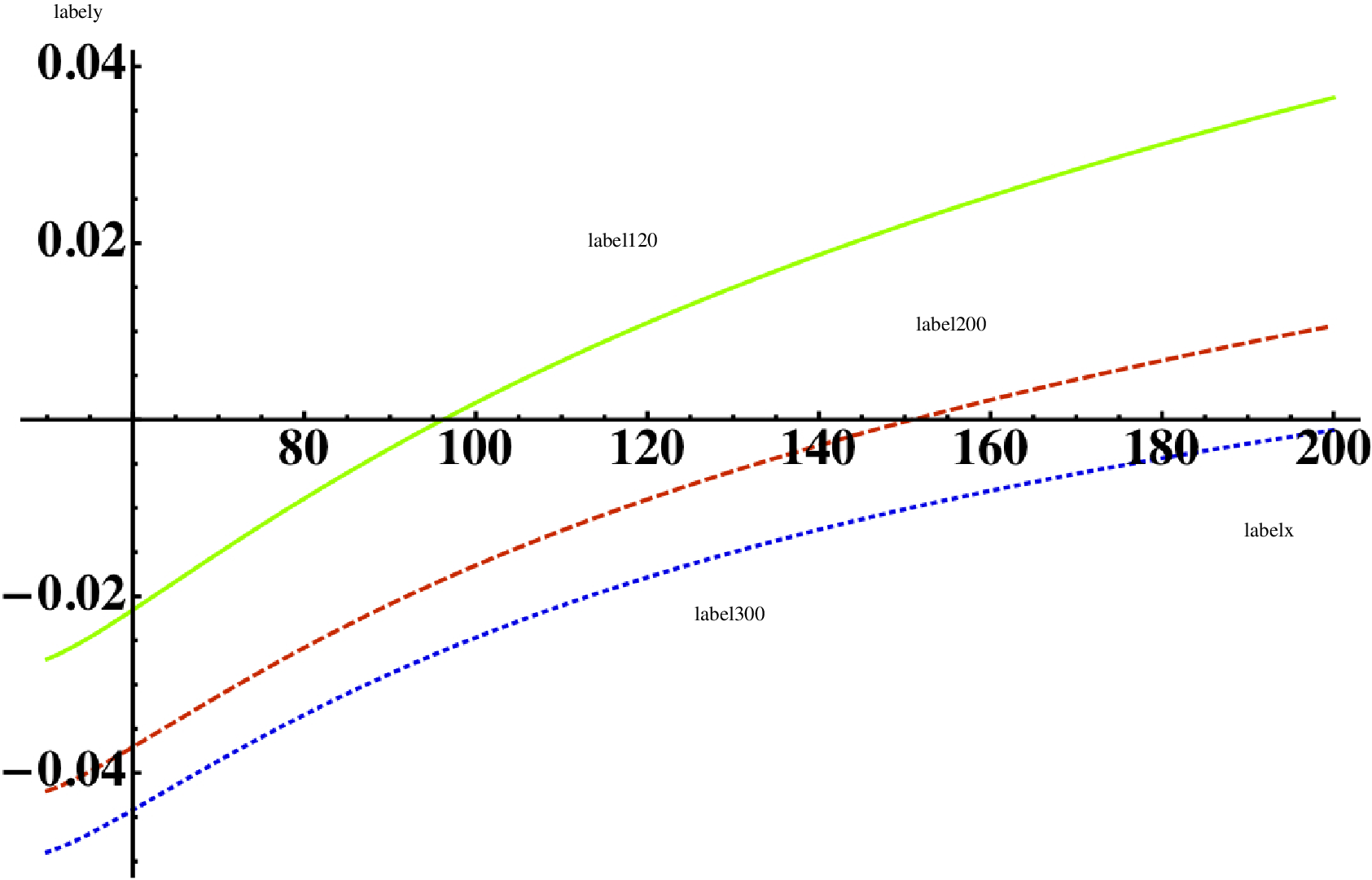}

\caption{Inert Higgses contribution to the S parameter as a function of $m_{\tilde{A}}$ and as a function of $m_{\tilde{S}}$ for various Higgs mass.}
\label{S_higgs}
\end{center}
\end{figure}

The Higgs doublets also give a  contribution to the S parameter \cite{Haber:1993wf}:
\begin{align} S&= \frac{1}{ \pi m_Z^2} [  B_{22}( m^2_Z, m^2_{\tilde{S}},m^2_{\tilde{A} })- B_{22}( m^2_Z, m^2_{\tilde{S}},m^2_{\tilde{A} })]
\end{align}
where:
\begin{align}
&B_{22}(q^2,m^2_1,m^2_2)=b_{22}(q^2,m^2_2,m^2_2)-b_{22}(0,m^2_1,m^2_2), \\  \newline
& b_{22}(q^2,m^2_2,m^2_1)=\frac{1}{4} ( \Delta+1) [m_1^2+m_2^2- \frac{1}{3} q^2]- \frac{1}{2}  \int_{0}^{1}  dx X \log{(X- i \epsilon)},  \\  \newline
& X= m_1^2 x+m_2^2 (1-x)- q^2 x(1-x),
\end{align}
and $ \Delta$ is the regulator of dimensional regularization defined by:
\begin{equation}
\Delta=\frac{2}{4-n}+\log( 4 \pi)+\gamma,
\end{equation}
$ n$ is the number of space-time dimensions and $ \gamma $ is the Euler's constant. The terms proportional to $ \Delta$ must clearly exactly cancel in the computation of the physical observables. In figure \ref{S_higgs} we show the contribution of the Higgs sector to the $S$ parameter for various Standard Model Higgs masses when the relationships \eqref{eq:msvsmhp} and \eqref{eq:mavsmsmh} hold.
\subsection{ Fermionic sector contributions}
\begin{figure}[h]
\begin{center}
\psfrag{labely}{\footnotesize $T$}
\psfrag{labelx}{\footnotesize $b$ (GeV)}
\psfrag{label11}{\footnotesize $1.1$ TeV}
\psfrag{label13}{\footnotesize $1.3$ TeV}
\psfrag{label15}{\footnotesize $1.5$ TeV}
\includegraphics[width=7.5cm]{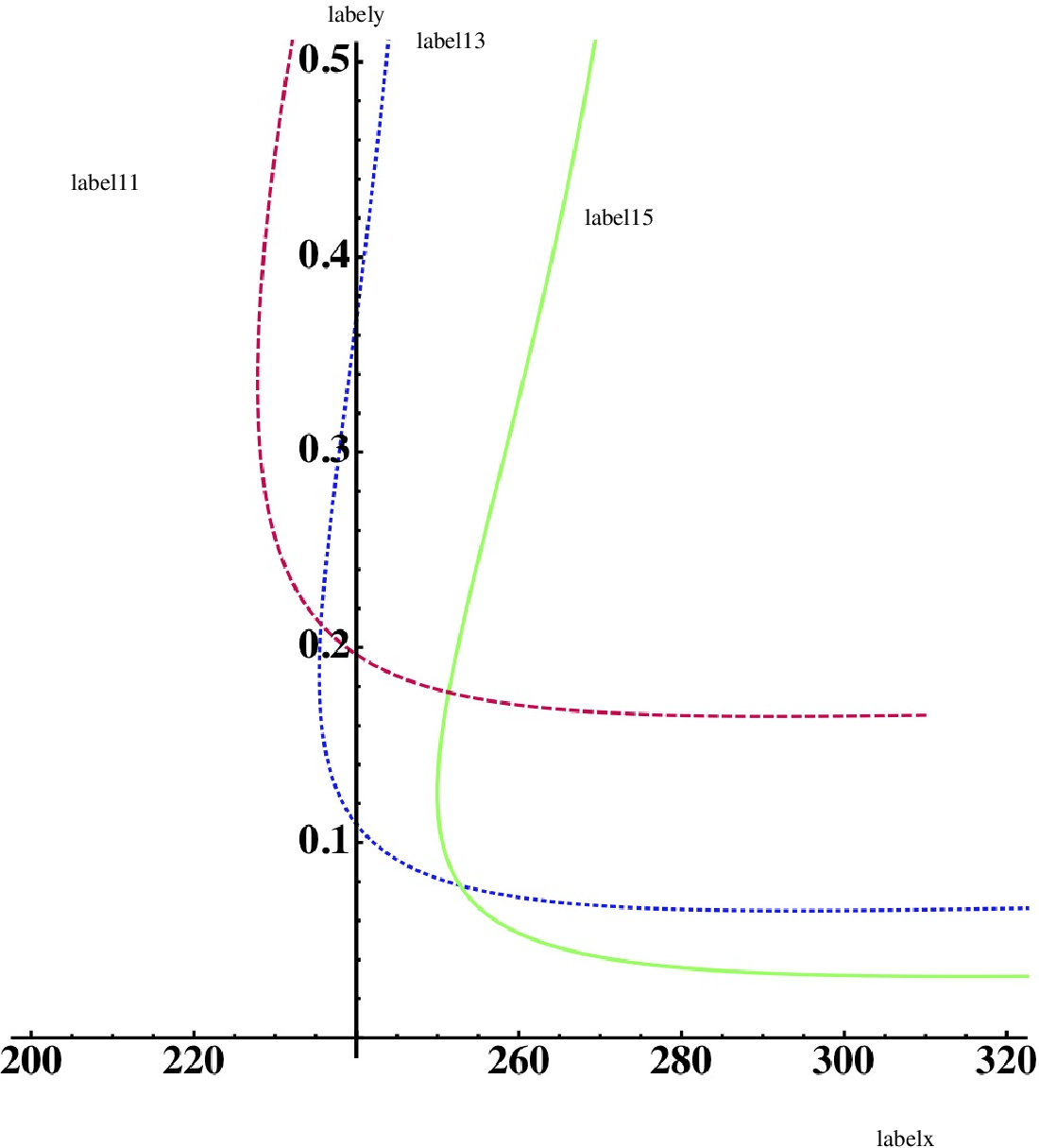}
\includegraphics[width=7.5cm]{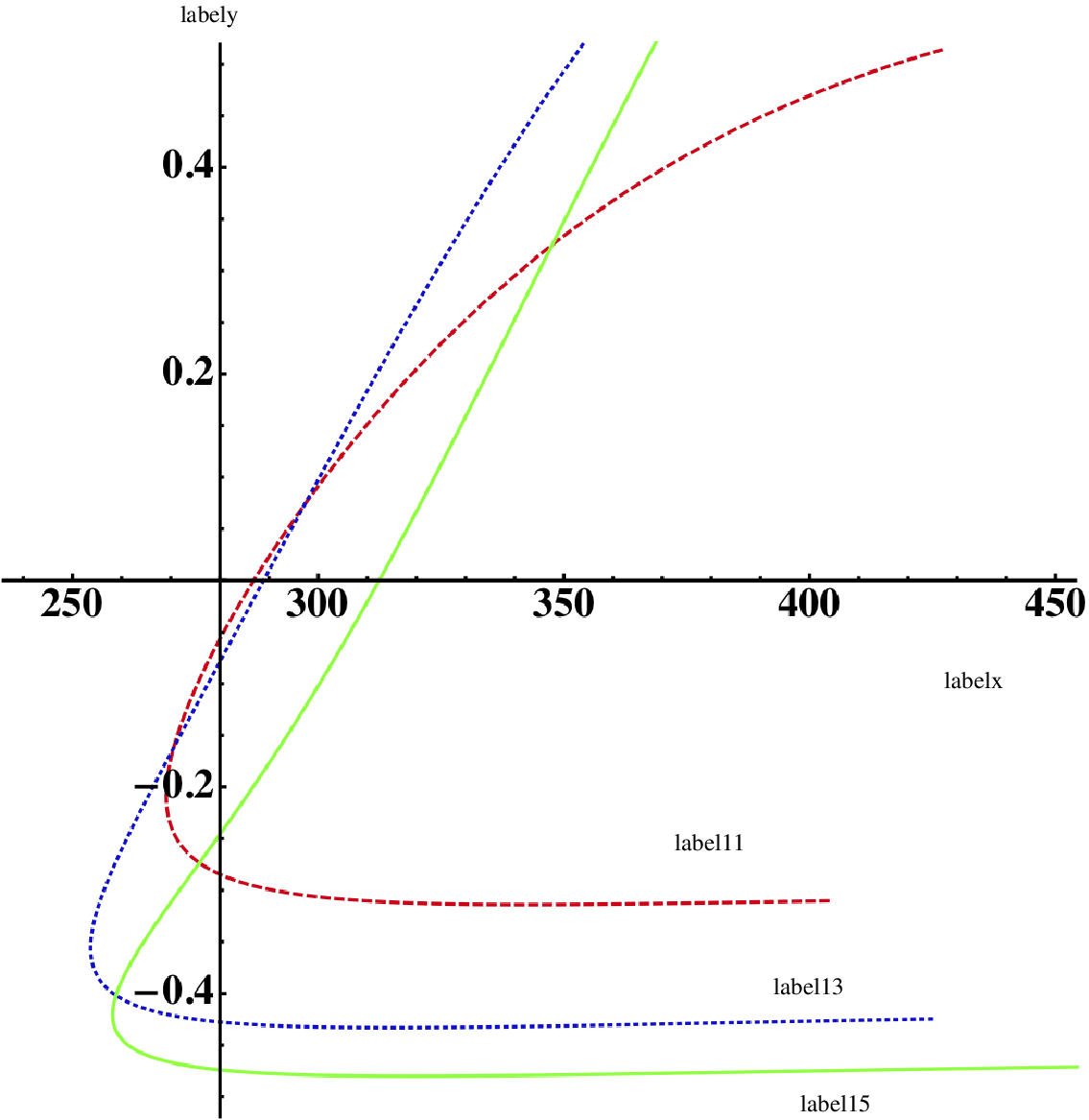}
\caption{$T$ parameter contribution from the top sector as a function of $b$ for various $m_U$ and with $f=700$ GeV. The plot on the left is for the $SU(6)_L \times SU(5)_R$ top sector with $m_D=1$ TeV while the plot on the right is for the $SU(6)$ top sector. }
\label{T_top}
\end{center}
\end{figure}

Loops of top partners, even if their contributions are parametrically suppressed  can lead to numerically significant contributions  to the $T$ parameter \cite{Gregoire:2003kr}.  In figure \ref{T_top} we show the contribution to $T$ for the $SU(6)$ and $SU(6)_L \times SU(5)_R$ top sectors  for $f=700 \text{GeV}$ and $M_{Q}=1 \text{TeV}$, as a function of $b$. We see that the contributions from the $SU(6)$ top sector are smaller due to the fact this the top Yukawa coupling was implemented in way that respect a custodial symmetry.

The light fermions will not contribution to the oblique parameter as their coupling to the Higgs sector is very small. On the other hand the Lagrangian eq.\eqref{eq:Llight}  contains vertices with one standard model fermion, the odd partner and the triplet $ \tilde{\Phi}$ :
\begin{equation}
 k  f_k  q_{\text{SM}}^\dagger \tilde{\Phi} \tilde{q} \;.
 \end{equation}
  This, at 
one loop through a box diagram, gives a  finite contribution to the four-fermion operator 
\begin{equation}
 q_{\text{SM}}^\dagger \sigma^{\mu}  q_{\text{SM}}  q_{\text{SM}}^\dagger \sigma_{\mu} q_{\text{SM}} ,
\end{equation}
as found in \cite{Hubisz:2005tx}. It leads to the constraint: $ k f_k \lesssim 1 $ TeV if we assume an universal $ k$\cite{Hubisz:2005tx}. 

\section{Dark Matter}
\label{sec:DM}
In this model the T-parity is exact, and the lightest T-odd particle (LPOP) will be stable and will contribute to dark matter abundance. The model contains a few T-odd particles with masses in the $\sim 100$ GeV rage that could be the LPOP, namely the odd triplet and the odd  Higgs doublet.  We will consider the case where $\tilde{S}$, the CP-even component of the inert Higgs doublet is the dark matter. In our model $\tilde{S}$ is always lighter than $\tilde{A}$, and we can choose it to be lighter than $\tilde{h}^+$. The relic abundance is completely determined by the parameters of the Higgs potential and was computed for various choices of parameters in \cite{Cirelli:2005uq,Dolle:2009fn,Hambye:2007vf,LopezHonorez:2006gr,LopezHonorez:2010tb}.   Using micrOMEGAs \cite{Belanger:2008sj} we calculated the relic abundance for parameters that are appropriate for our model, namely a nearly degenerate CP-even and charged T-odd states and masses of the Higgses doublet related as in  \eqref{eq:mavsmsmh}. We fix $m_h$ at $120$ GeV and vary $m_{\tilde{S}}$ and $\delta$ (see \eqref{eq:mhpmsdelta}). The coupling $\lambda_3$ ,which in our case control the coupling of two $\tilde{S}$'s to the Higgs and play an important role in the dark matter abundance calculation is taken to be $-2 \lambda_1=-m_h^2/v^2$. This is appropriate if $c_3$ is the larger than the other quartic couplings in the $h_A,h_B$ basis.

\begin{figure}[h]
\begin{center}
\psfrag{labelx}{$m_{\tilde{S}}$ (GeV)}
\psfrag{labely}{$\delta$}
\includegraphics[height=7cm]{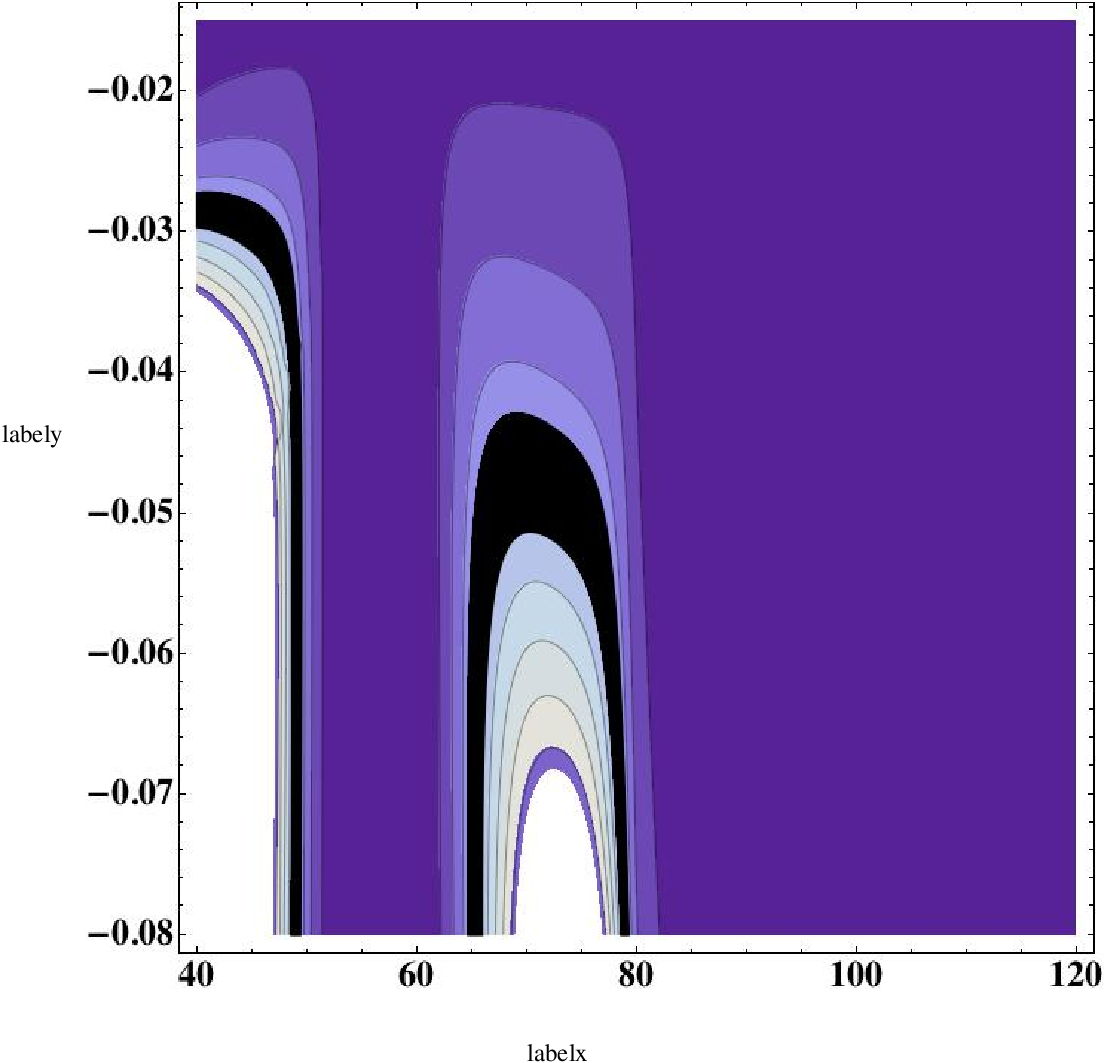}
\includegraphics[height=7cm]{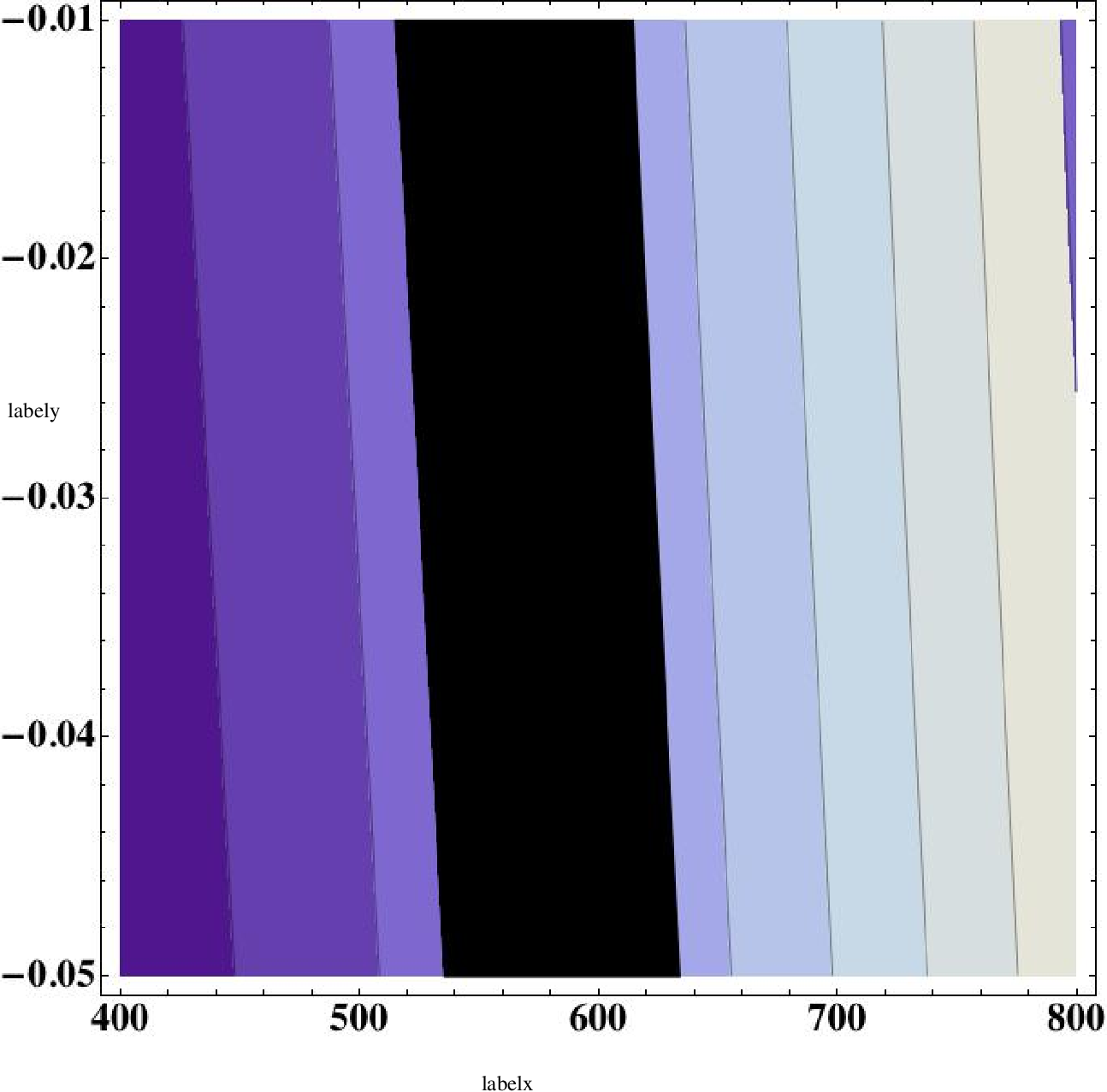}
\caption{Contour plots of $\Omega h^2$ as a function of $m_{\tilde{S}}$ and $\delta$. The region in black correspond  to the dark matter abundance measured by WMAP . The plot on the left shows a low $m_{\tilde{S}}$ region, while the plot on the right shows a high $m_{\tilde{S}}$ region, both have the Standard Model Higgs mass set at $120$ GeV.}
\label{fig:omega}
\end{center}
\end{figure}

If $\tilde{S}$ is heavier than roughly twice the $Z$ mass, allowing its annihilation into a pair of gauge bosons it will lead  to a dark matter abundance that is below the observed abundance for $m_{\tilde{S}}$ not too far  than this threshold. For light $m_{\tilde{S}}$, the only possible annihilation channel is to $b \bar{b}$ in which case the cross section is too small and the relic abundance is too large.  In between these two regions, there are typically regions where the relic abundance is correct.   Also, if the parameters are such that  $m_{\tilde{S}}$ can be near $m_h/2$, but below $m_W/2$, there in an additional enhancement of the annihilation cross section due to the possibility of exchanging an on-shell Higgs. As found also in \cite{Dolle:2009fn} this lead to a possible dark matter candidate of mass $\sim 70$ GeV with a Higgs mass at $120$ GeV. As a consequence of \eqref{eq:mavsmsmh}, this scenario would imply a light $m_{\tilde{A}}$, which seem hard to accommodate with the calculable $b$ values presented in figure \ref{fig:b}. Note however that other contribution to $b^2$ could be added. There is also a possibility of having a heavier dark matter candidate, with mass near $500$ GeV and a Higgs still around $120$ GeV. This scenario would require a slight fine-tuning to get the light Higgs. In figure (\ref{fig:omega}) we show contour plots that illustrate the allowed region for low and high $m_{\tilde{S}}$ masses.

Because $\tilde{S}$ and $\tilde{A}$ are split in mass, direct detection through a $Z$ exchange is small. Direct detection through a Higgs exchange could be large if $\left| \lambda_3 \right|$ is large. In our case, because $c_3$ is the dominant coupling this would mean a high Higgs mass. Also if $\tilde{S}$ and $\tilde{h}^+$ are degenerate enough, direct detection could proceed through inelastic scattering \cite{TuckerSmith:2001hy}. This would require a mass splitting of order $\sim 100$ keV, which correspond to $\delta \lesssim 0.001$.

\section{Phenomenology}
\label{sec:pheno}
The phenomenology of the $\sim$ TeV scale particles of this model is similar to other little Higgs models with T-parity \cite{Hubisz:2005tx,Belyaev:2006jh,Carena:2006jx}. There are odd and even top 'partners', the later being similar to that of littlest Higgs or other little Higgs model without T-parity and the former having signature potentially similar to the stop \cite{Meade:2006dw}. There are new odd gauge bosons, which would be produce in pair and also decay through a cascade leading to missing energy signatures. However, the cascade in this model will end-up in the Higgs sector with particles of masses of order $\sim 100$ GeV, as opposed to T-odd gauge bosons of $\sim$ TeV scale mass as is the case in the original littlest Higgs model with T-parity. 

The phenomenology of the odd doublet has been somewhat studied in the context of inert Higgs double models \cite{Cao:2007rm,Dolle:2009ft,Miao:2010rg} and can have signatures similar to neutralino and charginos. In particular, it is possible to have tri-lepton signatures and Tevatron bounds on chargino translate into a bound of $\sim 70 - 90$ GeV for $m_{\tilde{h}^+}$ in inert Higgs doublet models \cite{Dolle:2009fn}. However, note that since our charged and neutral CP-even Higgs are almost degenerate, the kinematic of the decay might be different than the one studied in \cite{Dolle:2009fn} and the bound might not apply in this case. 

Finally, the new odd triplet could provide another handle to distinguish this kind of model from other T-parity models and would be interesting to study in more details. 

\section{Conclusions}
\label{sec:conclusions}
Little Higgs models without T-parity suffer from important constraints on their parameter space due to deviations in electroweak precision tests. These constraints are eased significantly by the introduction of T-parity. However, T-parity tends to introduce new structure in the theory. On the one hand, the fermion sector of T-parity invariant models needs a larger global symmetry structure, and contain possible new flavor spurions that need to be close to the identity \cite{Hubisz:2005bd}. The second issue concerns the implementation of T-parity in the UV theory, which was the motivation for this work. 

The action of T-parity on the non-linear sigma model field of the littlest Higgs model was originally proposed to be of the form:
\begin{equation}
\label{eq:symLHST}
\Sigma \rightarrow \Omega \Sigma^\dagger \Omega \; \; \; \; \; \; \text{with }\Omega \in SO(5) \;.
\end{equation}
If $\Sigma$ is a composite of fermions, the corresponding symmetry in the UV theory is not straightforward to implement. The situation is in many ways similar to what happens in QCD. Looking only at the chiral Lagrangian, it seems easy to impose a $Z_2$ under which the pions and kaons are odd:
\begin{equation}
\label{eq:symqcd}
\Sigma_\text{QCD} \rightarrow \Sigma_\text{QCD}^\dagger \; \;.
\end{equation}
However, it is well known that this is not a symmetry of the QCD chiral Lagrangian. The kinetic term of the non-linear sigma model field is invariant, but there are terms, such as the $\pi^0 F \tilde{F}$ term or the WZW term that break it and allow coupling of one $\pi$ to two photons, or two $\pi$'s to three $K$'s. In fact, the high energy Lagrangian containing the quarks  and gluon has no symmetry that would lead to (\ref{eq:symqcd}) at low energy. 
Similarly, a symmetry of the form shown in (\ref{eq:symLHST}) can not be straightforwardly implemented in a strongly coupled UV completion such as the one of \cite{Katz:2003sn}. In this work, we examined the possibility of implementing T-parity in a different way, in the $SU(6)/Sp(6)$ little Higgs model. In this model the T-parity is an element of $Sp(6)$ and could easily be implemented as an exchange symmetry in a strongly coupled UV completion. We did not build an explicit UV completion as it would present similar challenges as in other little Higgs models, especially in the fermion sector. We instead focused on the description of the theory below $\sim 10$ TeV.

 One of the distinguishing feature of this model is the presence at low energy of an inert two Higgs doublet model. The neutral state CP even state  of the T-odd doublet is a potentially viable dark matter candidate, and would lead to interesting collider phenomenology, similar to gaugino production. The Higgs sector also respects an approximate custodial symmetry that makes the CP even neutral Higgs and the charged Higgs quite degenerate. It prevents this sectors from giving large contributions to the $T$ parameter and could lead to interesting effects at colliders. To give Yukawa couplings to fermions in a T-parity invariant way, we introduced a new $SU(2)$ gauge symmetry as well as two 'link' fields that break the resulting $SU(2)^3$ gauge symmetry down to the diagonal subgroup. The extra gauge field can be made heavy, but the link fields contain an extra T-odd scalar triplet which get a mass in the $\sim 100$ GeV range. The presence of such a triplet coming from link fields seem to be a generic feature of simple group little Higgs model with T-parity \cite{Pappadopulo:2010jx}. It could also be a dark matter candidate, and its phenomenology would be interesting to study further. We have also shown that this model could be consistent with electroweak precision data, and yield the correct dark matter abundance. 
 \section*{Acknowledgements}
 This work was in part support by the Natural Sciences and Engineering Research Council of Canada. TG and CF would like to thank the University of Edinburgh and the University of Sussex were part of this work was done.
  \bibliography{draft}
\bibliographystyle{JHEP}
\end{document}